\documentclass[useAMS,usenatbib]{mn2e}
\usepackage{epsfig}
\usepackage{subfigure}
\usepackage{amsmath,amssymb}
\usepackage{float}
\newcommand{\kms}{\>{\rm  km}\,{\rm  s}^{-1}}
\newcommand{\hMsol}{{\>h^{-1}\rm M}_\odot}
\newcommand{\hMpc}{{\>h^{-1}\rm  Mpc}} 
\newcommand{\hkpc}{{\>h^{-1}\rm kpc}}  
\newcommand{\Gyr}{\>{\rm Gyr}}

\title{The impact of environment on the dynamical structure of satellite systems}

\author[Faltenbacher]{
  A. Faltenbacher
  \thanks{E-mail:afaltenbacher@uwc.ac.za}
  \\  
  Physics Department,University   of   the   Western   Cape,  
  Cape   Town   7535,   South
  Africa\\
  Max-Planck-Institute      for      Astrophysics,
  Karl-Schwarzschild-Str.   1,  D-85741  Garching,   Germany\\  
  MPA/SHAO  Joint  Center  for  Astrophysical  Cosmology  at  Shanghai
  Astronomical Observatory, Nandan Road 80, Shanghai 200030, China}

\begin{document}
\date{\today}
\pagerange{\pageref{firstpage}--\pageref{lastpage}} \pubyear{0000}
\maketitle
\label{firstpage}
\begin{abstract}
  We examine the effects of  environment on the dynamical structure of
  satellite systems based on the Millennium--II Simulation.  Satellite
  halos are defined  as sub--halos within the virial  radius of a host
  halo. The  satellite sample is restricted to  those sub--halos which
  showed a  maximum circular  velocity above $30\kms$  at the  time of
  accretion.    Host    halo   masses   range    from   $10^{11}$   to
  $10^{14}\hMsol$.   We  compute  the  satellites'  average  accretion
  redshift, $z_{\rm acc}$, velocity dispersion, $\sigma$, and velocity
  anisotropy parameter,  $\beta$, utilising stacked  satellite samples
  of  equal mass  hosts  at similar  background  densities.  The  main
  results are: (1) On average  satellites within hosts in high density
  environments are accreted earlier ($\Delta z\approx0.1$) compared to
  their  counterparts  at  low   densities.   For  host  masses  above
  $5\times10^{13}\hMsol$ this trend weakens and may reverse for higher
  host  masses;  (2) The  velocity  dispersion  of  satellites in  low
  density  environments follows that  of the  host, i.e.   no velocity
  bias is observed for host halos at low densities independent of host
  mass.  However, for low mass  hosts in high density environments the
  velocity dispersion of the satellites can be up to $\sim30\%$ larger
  than that  of the  host halo, i.e.   the satellites  are dynamically
  hotter than their host  halos.  (3) The anisotropy parameter depends
  on host mass and environment.  Satellites of massive hosts show more
  radially  biased velocity  distributions.  Moreover  in  low density
  environments  satellites   have  more  radially   biased  velocities
  ($\Delta\beta\gtrsim0.1$)  compared to  their  counterparts in  high
  density environments. We believe that our approach allows to predict
  a similar behaviour for observed satellite galaxy systems.
 \end{abstract}
\section{Introduction}
\label{sec:intro}
The dependence of halo statistics on a second parameter in addition to
mass is now generally referred to as assembly bias.  Simple extensions
to     the     Press-Schechter     and    excursion     set     models
\citep{Press-Schechter-74,    Kaiser-84,    Bond-91,   Cole-Kaiser-89,
  Lacey-Cole-93,  Mo-White-96}  predict  the  clustering of  halos  to
depend on their  mass alone. However, \cite{Gao-Springel-White-05} and
various  subsequent studies  showed  that clustering  also depends  on
other  halo properties,  for example,  formation  time, concentration,
substructure  content, spin  and shape  \citep{Harker-06, Wechsler-06,
  Bett-07,   Gao-White-07,   Jing-Suto-Mo-07,  Maccio-07,   Wetzel-07,
  Angulo-Baugh-Lacey-08}.  Using a  mass filter in configuration space
rather than  in k-space \cite{Zentner-07}  demonstrated that excursion
set models  predict that halos  in denser environments do  form later,
independent  of halo  mass.  At  the high  mass end  this  agrees with
findings              from              N-body             simulations
\citep[e.g.,][]{Wechsler-02,Jing-Suto-Mo-07},  but is  opposite  to the
behaviour observed  at the low mass  end.  Based on  statistics of the
peaks within Gaussian random fluctuations, \cite{Dalal-08} argued that
the behaviour  for low  mass halos can  be understood if  cessation of
mass accretion is taken into account. A similar argument has been
proposed by \cite{Hahn-09}.\\

Several  other  studies  have  investigated  the  dependence  of  halo
formation   times   or,  similarly,   merger   rates  on   environment
\citep[e.g.,][]{Gottlober-Klypin-Kravtsov-01,             Gottlober-02,
  Sheth-Tormen-04,    Fakhouri-Ma-09,    Fakhouri-Ma-10,   Hahn-07}.
Although slightly  different density estimators are  employed, such as
over  density  in  a  sphere  or mark  correlation  functions,  it  is
generally    agreed    upon    that    halos   less    massive    than
$\sim10^{13}\hMsol$,  which  reside  in  high  density  regions,  form
earlier compared  to those with equal  mass but located  in less dense
regions.\\

In  a preceding  study we  discussed the  dependence of  the dynamical
structure     of     dark      matter     halos     on     environment
\citep{Faltenbacher-White-10}.  We   have  found  that   the  velocity
dispersion of  dark matter halos  residing in less  dense environments
shows a more  radially biased velocity structure compared  to halos of
the same mass in denser environments.  We now extend this study to the
dynamics of satellites within  host halos.  Satellites are selected on
the  basis that  our findings  can  be applied  to observed  satellite
galaxy systems.  The effects of environment on the dynamical structure
of  satellites  may  have  important implications  for  the  accretion
processes of the host galaxies.\\

The  outline of  the paper  is  as follows.   In \S~\ref{sec:meth}  we
introduce  the simulation  and  the halo  finding  procedure. We  then
explain how the  environment of host halos is  determined and describe
the  stacking procedure  to  compute average  properties of  satellite
populations.   \S~\ref{sec:results} presents  the  results: host  halo
properties   as    a   function   of   environment    are   shown   in
\S~\ref{sec:host}; main results are portrayed in \S~\ref{sec:s6} where
we discuss the accretion  history and dynamical structure of satellite
systems.   In \S~\ref{sec:res} we  investigate resolution  effects.  A
conclusion is given in \S~\ref{sec:conclusion}.
\begin{figure*}
  \epsfig{file=
    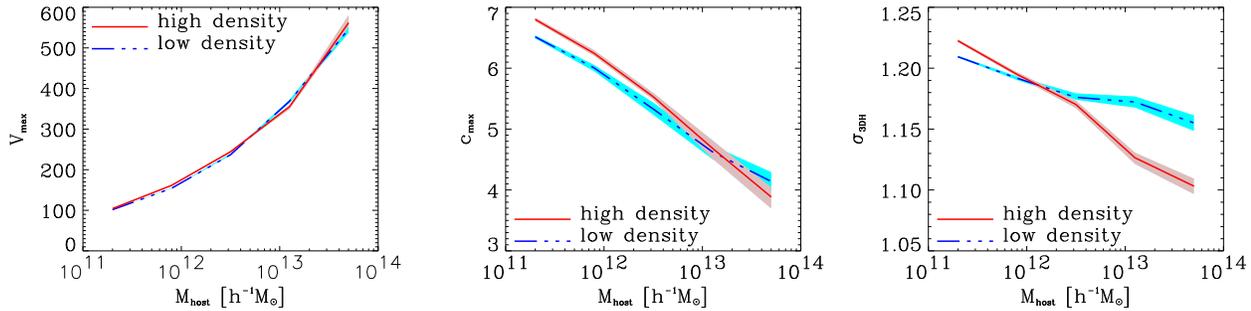,width=0.95\hsize }
  \caption{\label{fig:h3}   Average  host  halo   properties,  maximum
    circular velocity  ($V_{\rm max}$), concentration  ($c_{\rm max}$)
    and 3D velocity dispersion  ($\sigma_{\rm 3DH}$), as a function of
    halo  mass.   Red  (solid)  and blue  (dot-dashed)  lines  display
    results based on halos in the  upper and lower 33\% tails of their
    background  density distributions.   Halo  concentrations, $c_{\rm
      max}$, are defined as the ratio of the virial mass to the radius
    of the  maximum circular velocity.  Before  averaging the velocity
    dispersion  of  each  individual  halo  is scaled  by  its  virial
    velocity, $V_{\rm  vir}=(GM_{\rm vir}/R_{\rm vir})^{1\over2}$. The
    shaded regions  indicate the 1$\sigma$  confidence intervals based
    on a bootstrap resampling.}
\end{figure*}
\section{Methodology}
\label{sec:meth}
In  this section  we describe  the tools  for our  analysis.   A short
account on the Millennium--II Simulation is followed by the discussion
of the halo  finding procedure.  We also explain  the determination of
the  background density  and  the stacking  procedure  used to  derive
averaged properties of satellite systems in different environments.
\subsection{Simulation}
\label{sec:sim}
The Millennium--II Simulation \citep{Boylan-Kolchin-09a, Springel-05b}
adopted concordance values for the parameters of a flat $\Lambda$ cold
dark matter ($\Lambda$CDM) cosmological model, $\Omega_{\rm m} = 0.25$
and  $\Omega_{\rm  b} =  0.045$  for  the  current matter  and  baryon
densities,  $h =  0.73$ for  the  present dimensionless  value of  the
Hubble constant, $\sigma_8 = 0.9$  for the rms linear mass fluctuation
in a sphere  of radius $8\hMpc$ extrapolated  to $z = 0$, and  $n = 1$
for the slope of  the primordial fluctuation spectrum.  The simulation
followed $2160^3$ dark  matter particles from $z= 127$  to the present
day within a cubic region $100\hMpc$ on a side resulting in individual
particle masses of $6.9\times10^6\hMsol$.  The gravitational force had
a Plummer-equivalent comoving softening of $1\hkpc$.  We refer readers
to  \cite{Boylan-Kolchin-09a}  for more  detailed  description of  the
simulation.
\subsection{Halo sample and accretion times}
\label{sec:halo}
The halos  are found by a  two-step procedure.  In the  first step all
collapsed  halos with  at least  20 particles  are identified  using a
friends-of-friends (fof) group-finder with  linking parameter b = 0.2.
These objects will be referred to as fof--halos.  Then post-processing
with the substructure algorithm SUBFIND \citep{Springel-01} subdivides
each fof--halo  into a set  of self-bound sub--halos.   Sub-halos with
lest  than 20  particles are  discarded.   Here we  consider the  most
prominent  sub--halo   within  each   fof--halo  as  host   halo.   To
characterise  a  host halo  we  use:  the  maximal circular  velocity,
$V_{\rm  max}$,   which  peaks  at   a  radius,  $R_{\rm   max}$;  the
concentration, $c_{\rm  max} = R_{\rm vir}/R_{\rm  max}$, were $R_{\rm
  vir}$  denotes the  virial  radius,  i.e., the  radius  of a  sphere
centred at  the potential minimum of  the host which  comprises a mean
density  of 94 times  the critical  value.  For  a Navarro-Frenk-White
density profile \citep[][]{Navarro-Frenk-White-97}, $R_{\rm max}$ is a
factor of  $\sim2.1$ larger than  the scale radius $R_{\rm  s}$, which
has  conventionally been used  to compute  the concentration;  and the
three  dimensional velocity  dispersion, $\sigma_{\rm  3DH}$,  wich is
based on the  velocities of all host halo  particles after subtracting
their common bulk velocity from each of them.  The velocity dispersion
of  each individual  halo, $\sigma_{\rm  3DH}$, is  scaled  its virial
velocity $V_{\rm  vir}= (G M_{\rm  vir}/R_{\rm vir})^{1\over2}$, where
$M_{\rm  vir}$  is  the  virial  mass and  $G$  is  the  gravitational
constant.\\

Besides the {\it  host halo}, all other sub--halos  with more than 100
particles  and  located within  the  virial  radius  of the  host  are
referred  to as  {\it satellite  halos}.   We only  take into  account
sub--halos within  the virial radius of  the host halo  to prevent the
contribution of sub--halos within falsely linked `dumbbell' shaped fof
halos.  The  lower limit  of 100 particles  is imposed to  guarantee a
reliable determination  of the satellites'  velocities.  The accretion
redshift of a  satellite halo onto the host halo  is determined as the
redshift at which  it discontinues to be the  most prominent sub--halo
in  its own  fof--halo.   The time  elapsing  between two  consecutive
snapshots  is  roughly $0.3\Gyr$  which  determines  the accuracy  the
accretion times  used here.   We define the  redshift of  the snapshot
just prior  to accretion as {\it accretion  redshift}.  This typically
corresponds to the time when the halo acquires the maximum mass during
the  course  of its  evolution  \citep[cf.,][]{Guo-10}.  With  $V_{\rm
  max,acc}$ we  denote the  maximum circular velocity  of the  halo at
that time.  Following \cite{Kravtsov-Gnedin-Klypin-04} we include only
those  satellite  halos  in  our  analysis  with  $V_{\rm  max,acc}\ge
30\kms$.  This  restriction results in  a mean of few  1000 simulation
particles per  arriving satellite. Halos with  lower masses presumably
contain only negligible amounts  of stars. We abandon this requirement
only for a resolution study in \S~\ref{sec:res}.
\subsection{The background density field}
\label{sec:back}
The background density  field is computed based on  all sub-halos with
maximum  circular  velocities, $V_{\rm  max}$,  between $200\kms$  and
$300\kms$.   As Figure~\ref{fig:h3}  illustrates, this  velocity range
corresponds  to  halo  masses close  to  $M_*=6.15\times10^{12}\hMsol$
which is the  typical collapse mass at $z=0$  for the given cosmology.
Halos in this mass range provide a relatively unbiased sampling of the
overall  density field.  We  denote these  halos as  $V_*$--halos.  To
facilitate  the comparison  with density  estimates based  on observed
galaxies  we include all  sub--halos within  the given  velocity range
irrespective of  whether they are  host or satellite halos.   In total
there  are 2013 $V_*$  halos resulting  in a  sampling of  the density
field based on a point set  with a mean distance of slightly less than
$8\hMpc$.\\

The background density  for any given halo is  determined by the seven
nearest $V_*$ halos, each of  them smoothed with a smoothing kernel of
the form \citep{Monaghan-Lattanzio-85}
\begin{equation} 
  W(r;h)=\frac{8}{\pi h^3} \left\{
  \begin{array}{ll}
    1-6\left(\frac{r}{h}\right)^2 + 6\left(\frac{r}{h}\right)^3, &
    0\le\frac{r}{h}\le\frac{1}{2} ,\\
    2\left(1-\frac{r}{h}\right)^3, & \frac{1}{2}<\frac{r}{h}\le 1 ,\\
    0 , & \frac{r}{h}>1 .
  \end{array}
  \right.  
\end{equation}
As in \cite{Springel-Yoshida-White-01}  we define the smoothing kernel
on the interval $[0,h]$ and not  on $[0,2h]$ as frequently done in the
literature.  The  background density at  a given halo location  is the
sum of the  contributions of the seven nearest  neighbours. Using this
recipe, a {\it background density} is assigned to each host halo which
we also address  as {\it environment} of the  host.  In the subsequent
analysis  the host  halos are  ordered according  to  their background
density and average  properties of the upper and  lower 33\% tails are
determined separately.  We also  investigated the outcome based on the
upper and lower 20\% tails  and found very similar results. To improve
statistics we choose the larger background density intervals.
\subsection{The stacking procedure}
\label{sec:stack}
Besides deriving the average accretion redshift, $z_{\rm acc}$, of the
satellite populations as  a function of host mass  and environment, we
aim to  study differences in  their average dynamical  properties.  To
achieve reasonable statistical significance individual host halos with
similar masses and background densities are stacked.  For that purpose
we subtract  the bulk  velocity of the  host halo from  the individual
satellite  velocities and  scale them  by the  virial velocity  of the
host, $V_{\rm  vir}$. The scaling is  done to compensate  for the host
mass variation within the individual mass bins.\\

The scaled velocities of satellites belonging to host halos of a given
mass at a given background density are used to compute the mean radial
velocity,  $v_{\rm rad}$, the  three dimensional  velocity dispersion,
$\sigma_{\rm 3D}$, its  radial and tangential components, $\sigma_{\rm
  rad}$  and $\sigma_{\rm  tan}$, and  based on  these  quantities the
velocity   anisotropy  parameter,   $\beta=1   -  0.5\,   (\sigma_{\rm
  tan}^2/\sigma_{\rm  rad}^2$).  In  addition we  compute  the average
number of  satellites, $N_{\rm sat}$, as  a function of  host mass and
environment as  well as the  average ratio between the  actual maximal
circular  velocity  and  that  one  at the  time  of  the  satellites'
accretion, $V_{\rm max}/V_{\rm max,acc}$.
\section{Results}
\label{sec:results}
Before we  discuss the environment  dependence of accretion  times and
dynamical  structure of  satellite  systems we  review  the impact  of
environment  on  the  host  halo  properties  themselves.   Resolution
effects are discussed in the final paragraph.
\subsection{Host halo properties as a function of environment}
\label{sec:host}
The  left  panel  of  Figure~\ref{fig:h3} shows  the  average  maximum
circular  velocity, $V_{\rm  max}$, of  host  halos as  a function  of
mass.  Red  (solid) and  blue  (dot-dashed)  lines,  as with  all  the
remaining figures, show the result  for the halos within the upper and
lower   33\%   tails   of   the  background   density   distributions.
Subsequently, we will  address the two samples as  {\it high} and {\it
  low} density  samples. The figure indicates that  the average values
of $V_{\rm max}$ are independent of environment.

The  middle panel of  Figure~\ref{fig:h3} displays  the concentration,
$c_{\rm  max}$,  as  a  function  of host  mass  and  environment.  In
agreement  with previous  studies we  find  low mass  halos with  high
background  densities to be  more concentrated  compared to  their low
density  counterparts  of the  same  mass.  Again,  this behaviour  is
reversed     for     halo     masses     above     $\sim10^{13}\hMsol$
\citep[cf.,][]{Wechsler-06,  Gao-White-07, Jing-Suto-Mo-07, Maccio-07,
  Wetzel-07, Angulo-Baugh-Lacey-08, Faltenbacher-White-10}.

The  right  hand  panel   of  Figure~\ref{fig:h3}  depicts  the  three
dimensional velocity dispersion of the host halos, $\sigma_{\rm 3DH}$,
as a function of mass and environment.  Low mass halos in high density
regions   have   higher  velocity   dispersions   compared  to   their
counterparts in  low density regions.  This behaviour is  reversed for
higher masses which is similar to what is seen for the concentrations.
The analog  behaviour of concentration  and velocity dispersion  is in
agreement   with    predictions   based   on    the   Jeans   equation
\citep[e.g.,][]{Faltenbacher-Mathews-07}.   However,   we  notice  the
crossing  takes  place  at  some   lower  masses  than  seen  for  the
concentrations.
\begin{figure*}
  \epsfig{file=
    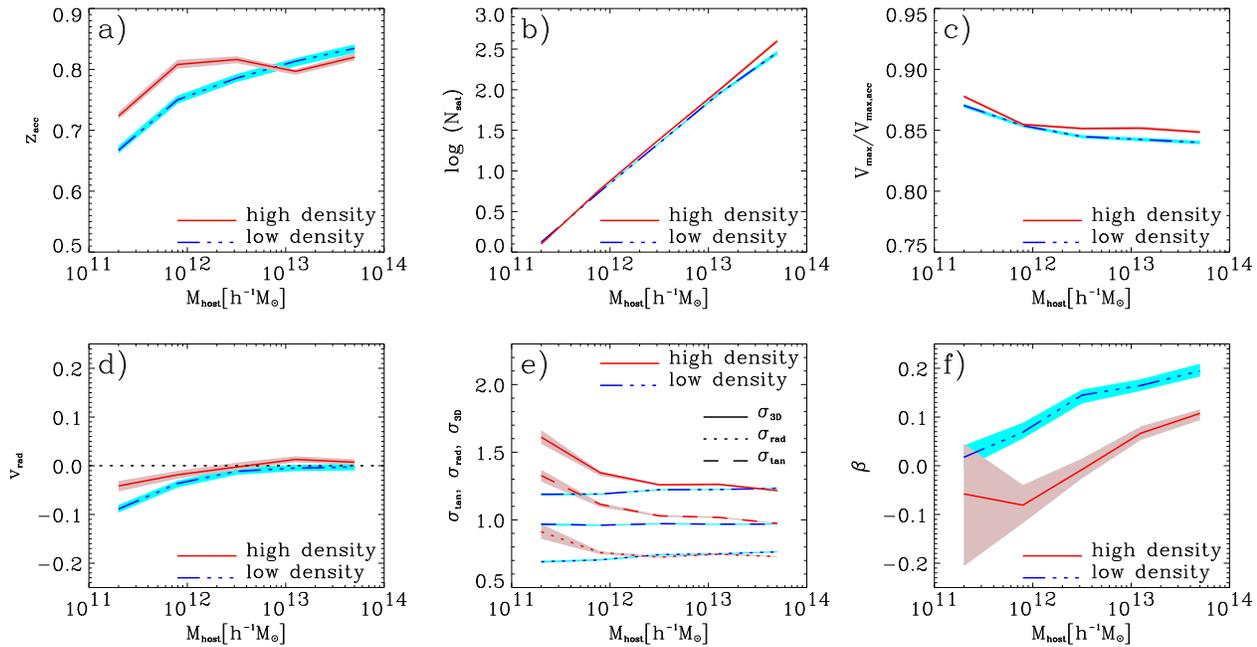,width=0.95\hsize }
  \caption{\label{fig:s6}  Average sub-halo halo  properties, redshift
    of accretion  ($z_{\rm acc}$), number  of sub-halos per  host halo
    ($N_{\rm  sat}$),  the  ratio  of  the  current  maximum  circular
    velocity  and  its  values  at  the  time  of  accretion  ($V_{\rm
      max}/V_{\rm max,acc}$), the radial velocity ($v_{\rm rad}$), the
    three dimensional velocity  dispersion ($\sigma_{\rm 3D}$) and the
    velocity anisotropy $\beta$  as a function of the  host halo mass.
    Red (solid)  and blue (dot-dashed) lines display  results based on
    host halos in  the upper and lower 33\%  tails of their background
    density distributions.  (By analogy, the upper and  lower lines of
    the pairs of  dashed and dotted lines correspond  to results based
    on  host  halos  in  the  upper  and lower  33\%  tails  of  their
    background density distributions.)   Before averaging the velocity
    of each  individual sub-halo is  scaled by the virial  velocity of
    its  host  halo.   The   shaded  regions  indicate  the  1$\sigma$
    confidence intervals based on a bootstrap resampling.}
\end{figure*}
\subsection{Accretion history and dynamical structure of satellite
  systems}
\label{sec:s6}
The  upper  left panel  of  Figure~\ref{fig:s6}  displays the  average
accretion redshift of satellites within host halos at high (red, solid
line) and low (blue, dot-dashed line) densities as a function of mass.
For      hosts     below      the      typical     collapse      mass,
$M_*=6.15\times10^{12}\hMsol$, satellites in  high density regions are
accreted earlier  compared to satellites  in low density  regions. For
host masses  above $M_*$ the  difference becomes marginal or  may even
reverse.  The  dependence of satellite accretion  times on environment
is  similar  to  the  observed  assembly  bias  for  halos  themselves
\citep[e.g.,][]{Gao-Springel-White-05}.  A similar correlation between
host halo mass and satellite accretion redshift has also been reported
in  a study  by  \cite{Lagos-Padilla-Cora-09}. This  behaviour can  be
explained by the fact that as the host mass increases the mean mass of
the  accreted satellites  increases as  well and  therefore  makes the
satellite population as a whole more resistant against tidal forces.

At  the first  glance  our  findings seem  the  contradict results  by
\cite{Fakhouri-Ma-09}. They find a modest increase of the merger rates
in  high   density  environments   at  present  time.    However,  the
measurement of the mean  accretion redshift of surviving satellites is
a convolution of the merger rate (over an extended period of time) and
the  survival rate  which may  depend on  environment as  the velocity
structure does.   In particular we  will see below that  satellites in
high  density environments show  a more  tangentially biased  velocity
structure which  prevents rapid destruction by the  hosts' tidal field
as it  is the case for  more radial orbits.  We  also checked, whether
the  satellites' masses  at the  time  of accretion  are dependent  on
environment.  It  turned out that this  is not the  case. Therefore, a
dependence of the  satellite masses on environment can  be excluded as
explanation  of  the  accretion  time  differences  of  the  surviving
satellite population.

The mean number of satellites for  the given host halo masses does not
depend on  environment as the  overlapping graphs in the  upper middle
panel of  Figure~\ref{fig:s6} indicate.  As  shown in the  upper right
panel the average  decrease of the maximum circular  velocity is about
15\% which is slightly reduced  for low mass hosts.  Consequently, the
central  profiles of  satellite halos  remain fairly  intact  on their
orbits within the  potential well of the host. This  behaviour is in a
agreement   with  the   study  by   \cite{Kazantzidis-04},   see  also
\cite{Boylan-Kolchin-09b}.

The lower  left panel of  Figure~\ref{fig:s6} depicts the  mean radial
velocity  of satellites as a function of  host mass  and environment.
All  velocities are scaled  by the  corresponding virial  velocity.  A
radial velocity close to zero  indicates that on average the satellite
distribution  is  static  which  is  observed  for  host  halos  above
$10^{12}\hMsol$.   Negative  values  correspond to  an  net infall or
some other excess of inward moving satellites.

\begin{figure}
  \begin{center}
    \epsfig{file= 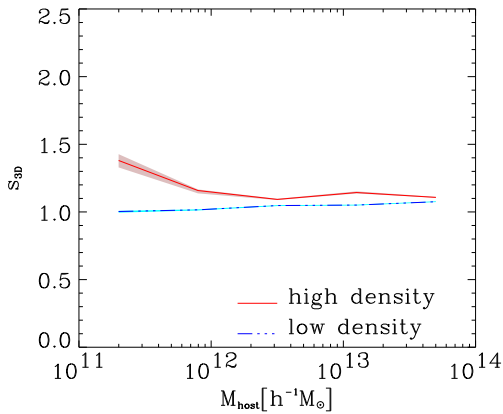,width=0.85\hsize }
  \end{center}
  \caption{\label{fig:s1} Satellite velocity  dispersion scaled by the
    host halo velocity dispersion, $s_{\rm 3D}$, as a function of host
    halo mass.  Except  for the scaling this figure  repeats the solid
    and  dot-dashed   lines  in  the  velocity   dispersion  panel  of
    Fig.~\ref{fig:s6}.}
\end{figure}

The lower  middle panel in Figure~\ref{fig:s6} displays  the 3D (solid
and  dot-dashed lines)  velocity  dispersion of  satellites and  their
radial (dotted lines) and the  tangential (dashed lines) as a function
of host  mass and environment.  All satellite  velocities are computed
relative  to the  bulk velocity  of the  host halo  and scaled  by the
corresponding virial velocity before  averaging.  We find a remarkable
rise in  the dispersions of the  satellites in low mass  hosts at high
densities.  These  satellites show  velocities 30\% larger  than their
counterparts in low density  environments. This behaviour is discussed
more thoroughly below in the context of Figure~\ref{fig:s1}.

The lower  right panel of Figure~\ref{fig:s6}  presents the anisotropy
parameter  of  satellite galaxies  as  a  function  of host  mass  and
environment. We find a  profound difference in the dynamical structure
of satellite systems in high density environments as compared to their
counterparts   at  low  background   densities  ($\Delta\beta\ge0.1$).
Independent  of host  mass, the  satellite velocities  in  low density
environments  are more  radially  biased than  those  in high  density
regions.  Besides  the dependence  on  environment  there  is also  an
obvious  trend  with  host  mass.   Independent  of  environment,  the
velocity structure  of satellites is more radially  biased within more
massive hosts.

Host  halos  in  a  low  density  environments  dominate  the  ambient
gravitational field  more strongly than their  equal mass counterparts
at  high  densities.  The  gravitational  field  lines point  radially
towards  the host  and  so  does the  acceleration  exerted on  future
satellites.  The dynamical structure of the satellites within the host
halo  is a  result  of the  fairly  radial inflow.   In contrast,  the
gravitational field  in high density regions is  more complex.  Before
satellites are  accreted onto a  host they also experience  non radial
accelerations  from other  massive  haloes nearby.   As  a result  the
velocities  of satellites  show  a larger  non-radial component.   The
dependence of $\beta$ on mass may be interpreted in the same way.  The
more massive  the host the more radial  the gravitational acceleration
of future satellites.  Which results in the more radially biased
velocity dispersions of the satellites within high mass hosts.\\

Figure~\ref{fig:s1} further explores  the large velocity dispersion of
satellites in  low mass hosts  at high background densities.   Here we
display  the  three--dimensional velocity  dispersion  of the  stacked
satellite  populations  scaled   by  the  measured  three  dimensional
velocity   dispersions   of   the   host   halos  and   not,   as   in
Figure~\ref{fig:s6},  by the  virial velocities.   Thus,  the velocity
dispersion  of the  satellites are  directly compared  to that  of the
host.   We find that  satellites in  host halos  at low  densities and
satellites in high mass  hosts (independent of environment) don't show
velocity   bias.     This   is   in   agreement    with   results   in
\cite{Faltenbacher-Diemand-06}  and \cite{Lau-Kravtsov-Nagai-09}.  The
velocity  dispersion  of   satellites  in  high  density  environments
deviates    from   this    behaviour.    For    host    masses   below
$\sim10^{12}\hMsol$ the  satellites are  hotter than the  overall dark
matter component.

This effect can not be explained by more recent accretion times since,
in  this case,  the low  density  satellites should  show even  larger
dispersions.         Recently,        \cite{Wang-Mo-Jing-07}       and
\cite{Fakhouri-Ma-09}   reported   on   dynamically   hooter   ambient
environments for halos in high density regions.  In addition, low mass
hosts  in  dense  environments   are  located  next  to  more  massive
halos. These halos expel a substantial fraction of their sub-halos out
to large radii  \citep[e.g.,][]{Gill-04, Ludlow-09}.  Thus, satellites
which merge onto or pass through  the low mass host systems are hotter
compared to  their counterparts at low densities.   It seems plausible
that  some  fraction of  satellites  within  low  mass hosts  at  high
densities do  originally not  belong to the  Lagrangean volume  of the
host  halo instead  they are  accelerated by  more  massive structures
nearby.  These satellites are dynamically hotter then the host itself.
In  the  following discussion  we  address  these  satellites as  {\it
  interlopers}. 
\begin{figure*}
  \epsfig{file=
    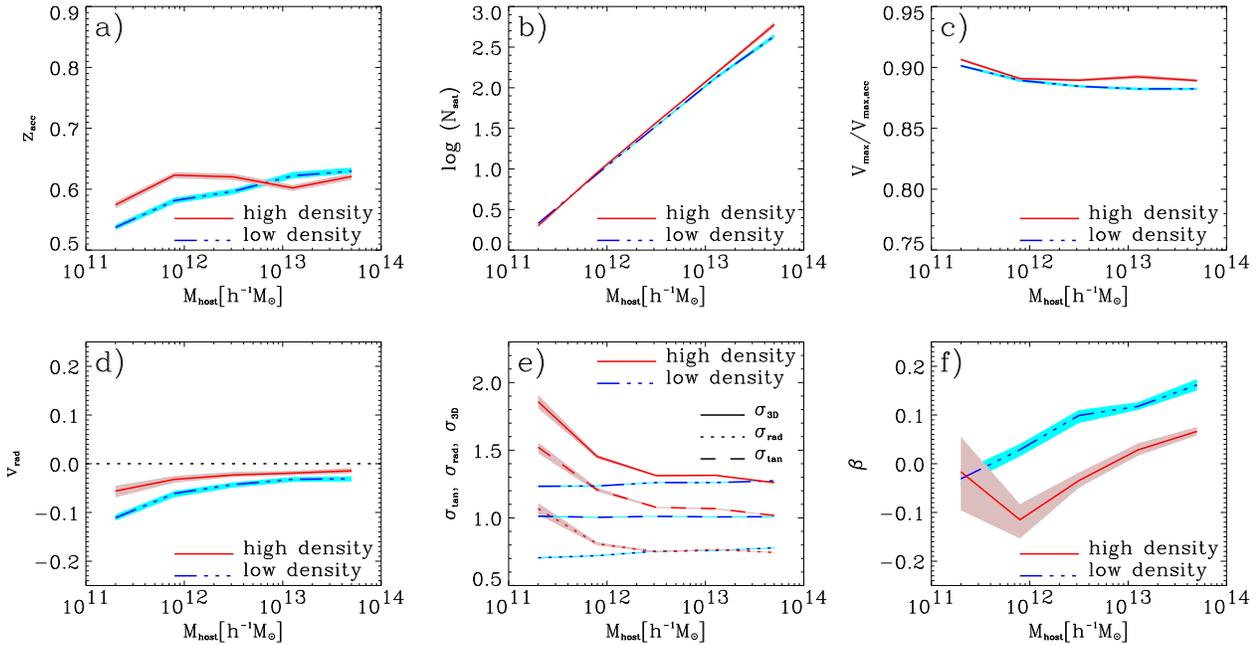,width=0.95\hsize}
  \caption{\label{fig:b6}  Same quantities  as  in Figure~\ref{fig:s6}
    but based on  all satellites, i.e. without any  restriction on the
    maximum circular velocity at the time of accretion.}
\end{figure*}
\subsection{Abandoning the mass barrier for entering satellites}
\label{sec:res}
Figure~\ref{fig:b6}    iterates     the    analysis    presented    in
Figure~\ref{fig:s6} with  the only difference that  the restriction on
the  maximum  circular velocity  at  the  time  of accretion,  $V_{\rm
  max,acc}\ge 30\kms$,  is abandoned.  The outcomes  presented in this
section should not be considered as  results per se. They are shown to
point out  the impact of the  $30\kms$ restriction. In  this case, all
satellites which can be detected at present, i.e.  all sub--structures
which have more than 100 particles at z=0, contribute to the averaging
process. If a  sub--halo falls below the 100  particle limit we assume
it as tidally {\it dissolved}.   Since many satellites have low masses
at the  time of accretion tidal  dissolution is more  prominent in the
unrestricted sample.  This is the key difference between the satellite
sample with and without restriction.   In the following we discuss the
individual  panels  in Figure~\ref{fig:b6}  and  compare  them to  the
corresponding  panels in  Figure~\ref{fig:s6}:\\{\bf  a)} The  average
accretion redshift,  $z_{\rm acc }$,  is reduced for  the unrestricted
sample.   This is expected  since small  mass satellites  have shorter
survival times.   Thus, a  large fraction of  early accreted  low mass
systems   is  missing.    However,  the   dependence   on  environment
persists;\\{\bf b)}  The average number  of satellites increases  by a
factor of  $\lesssim 2$.  This  is a consequence  of the shape  of the
(sub-)halo  mass  function  \citep{Boylan-Kolchin-09b};\\{\bf c)}  The
average fraction between current maximum circular velocity and that at
the  time of  accretion, $V_{\rm  max}/V_{\rm max,acc}$  is increased.
This  behaviour reflects the  low average  accretion redshift  for the
unrestricted   satellite   sample;\\{\bf   d)}  The   average   radial
velocities, $v_{\rm rad}$, are negative for all host masses. This is a
consequence  of  the  lower  mass  limit  of  100  particles  for  the
satellites.   As demonstrated in  \cite{Faltenbacher-Mathews-07} small
mass satellites are most likely  to fall below the resolution limit at
their  peri-centre  passage  causing  an  excess  of  negative  radial
velocities or inward moving satellites;\\{\bf e)} The average velocity
dispersions,  $\sigma_{\rm 3D}$,  $\sigma_{\rm rad}$  and $\sigma_{\rm
  tan}$,  increase  by  a  few  per  cent,  independent  of  mass  and
environment.  In addition there is  a substantial rise in the velocity
dispersions of  satellites in  low mass hosts  at high  densities. The
former  effect  can  be  explained  by  the  lower  average  accretion
redshifts.  More specifically, the  increase of velocity dispersion is
a result of the tidal dissolution of earlier accreted satellites which
have on  average lower velocities.   After their dissolution  the slow
moving satellites do not  contribute to the overall dispersion anymore
which causes a positive bias  of the velocity dispersion.  This effect
is  more noticeable for  the satellite  sample without  restriction of
$V_{\rm  max,acc}$ since  there tidal  dissolution is  more prominent.
The increase  of the velocity dispersion  in low mass  systems at high
density is presumably an enhanced contribution of low mass interlopers
in  high  density  environments;\\{\bf  f)}  The  velocity  anisotropy
parameter,  $\beta$,  decreases by  $\sim0.05$,  i.e.  the  tangential
velocity  component of the  satellites is  slightly increased  for the
sample without restriction on $V_{\rm max,acc}$.  In addition, we find
a distinct upward turn for  low mass hosts in dense environments.  The
former  is due  to  the fact  that  satellites on  more radial  orbits
penetrate deeper  into the potential well of  the host.  Consequently,
they  are more  strongly exposed  to  tidal forces  and get  dissolved
faster.  The  remaining satellite population is biased  towards a more
prominent  tangential velocity  component.  The  upward trend  for low
mass  hosts  at  high  densities  may  be  explained  by  interlopers.
Dynamically, they  are not strongly  coupled with the  host.  Assuming
random  motions of  the interlopers  one expects  $\beta=0$  which, in
deed, is nearly approached.

The comparison above highlights  the fact that satellites are modelled
by dark matter particles  of a finite mass ($6.9\times10^6\hMsol$) and
consequently  can not  be traced  further or  are lost  when  they get
tidally striped  below a  given `trustworthy' mass.  Here we  set this
mass  to  be  $6.9\times10^8\hMsol$  corresponding to  100  particles.
Satellites  which get  tidally striped  below this  limit 'disappear',
i.e., they  do not  contribute to the  average values  (like accretion
redshift,  velocity  dispersion etc.)  for  the satellite  populations
anymore.  This causes a bias towards later accreted satellites and the
consecutive effects.   In particular  the large number  of approaching
satellites with masses only slightly above the mass cut contributes to
that bias.  However, if  the restriction $V_{\rm max,acc}\ge30\kms$ is
imposed these effects  are not apparent.  In that  case the satellites
comprise on average  a few $1000$ particles at  the time of accretion.
Even substantial tidal stripping does not push the satellite below the
lower  particle limit,  i.e.   all arriving  satellites have  extended
survival    times.     Therefore,    the    results    presented    in
Figure~\ref{fig:s6} describe physical phenomena  and are not caused by
the finite masses of the simulation particles.
\section{Conclusion}
\label{sec:conclusion}
Taking  advantage of  the  superior resolution  of the  Millennium--II
Simulation  we  determine  the  impact  of  assembly  bias  on  galaxy
systems. We conclude with a recapitulation of the main results:\\ {\bf
  1)}  The average  accretion redshift  depends on  the host  mass and
environment.   For  host masses  of  $10^{11}\hMsol$  we find  $z_{\rm
  acc}\approx0.7$ which increases for host halos above $10^{13}\hMsol$
to  $z_{\rm  acc}\approx0.8$.    The  average  accretion  redshift  of
satellites  halos  in high  density  environments  is larger,  $\Delta
z\approx0.1$, compared  to that of  their counterparts in  low density
regions.  The  dependence of satellite accretion  times on environment
is  most  likely  caused  by  the more  tangentially  biased  velocity
structure (see  point 3) of  satellites in high density  regions which
makes  them more  resistant to  the  hosts' tidal  field. In  contrast
satellites  in  low density  environments  show  more radially  biased
velocity structures. As a  consequence total tidal dissolution of this
satellites  may  happen  faster  causing  a  lower  average  accretion
redshift of  the surviving satellite population.\\{\bf  2)} Host halos
above  $10^{12}\hMsol$  show the  same  velocity  dispersion as  their
satellite populations. This remains valid  for lower mass halos in low
density  environments.  However, the  dispersion of  satellites within
low  mass  hosts  at  high  densities  exceed that  of  the  hosts  by
$\sim30\%$. Which may be interpreted as contamination by high velocity
interlopers.\\ {\bf 3)} The  velocity anisotropy of satellite halos is
correlated with  the mass  of the host  halo. More massive  hosts show
more  radially biased  satellite velocities.   In addition  we  find a
strong   dependence  of  the   velocity  anisotropy   on  environment.
Satellite  velocities   of  host   halos  residing  in   high  density
environments tend do be less radially biased than those in low density
environments.\\

Our findings  show that the internal dynamical  structure of satellite
systems is correlated with  environment.  We believe that our approach
allows to make similar  predictions for real satellite galaxy systems.
Thus the dynamical structure of satellite galaxy systems should depend
on  environment.  Most  notably satellites  of host  galaxies  in less
dense  environments show  a more  radially biased  velocity structure.
This behaviour may have implications  for the accretion process of the
host galaxies.

\section*{Acknowledgements}
We would like to thank the anonymous referee for useful remarks, Simon
White  and Mike  Boylan-Kolchin  for insightful  comments and  Russell
Johnston for careful revisions which  helped to improve the paper.  It
is  a   pleasure  to   acknowledge  the  South   African  Astronomical
Observatory  or   kind  hospitality.   The   Millennium-II  Simulation
databases used in this paper  and the web application providing online
access  to them  were constructed  as part  of the  activities  of the
German Astrophysical Virtual Observatory.  Data for halos and galaxies
are publicly available at http://www.mpa-garching.mpg.de/millennium.
%

\label{lastpage}
%
%
\end{document}